\documentclass[%
 reprint,
 amsmath,amssymb,
 aps,
]{revtex4-2}

\usepackage{graphicx}
\usepackage{dcolumn}
\usepackage{bm}

\usepackage[margin=0.7in]{geometry}
\usepackage{mathptmx}
\usepackage{graphicx}
\usepackage{float}
\usepackage{subcaption} 
\usepackage{amsmath}
\usepackage{amssymb}
\usepackage{mathtools, nccmath}
\usepackage{fancyhdr}
\usepackage{slashed}
\usepackage{float}
\begin{document}

\title{Classical Stochasticity Using Quantum Computers}
\author{Diego Campos} \email{diego.camposmendez@uleth.ca,} \author{Narasimha Reddy Gosala}\email{narasimha.gosala@uleth.ca} \author{ Arundhati Dasgupta}\email{arundhati.dasgupta@uleth.ca}.
\affiliation{Physics and Astronomy,\\ University of Lethbridge,  4401 University Drive, \\ Lethbridge T1K 3M4, CANADA.}


\begin{abstract}
\noindent We demonstrate that quantum algorithms can be used to model classical stochastic simulations as the measurement process is inherently random. To illustrate, we solve the classical Lorenz system with stochastic behavior using a Python random number generator. We compare the classical stochastic Lorenz system with the measured output of the quantum algorithm.
\end{abstract}
\maketitle
Classical stochastic systems such as thermal gases, atmosphere, population dynamics, and financial chaos are modeled using random numbers. Monte-Carlo simulations of these systems are numerically obtained using computers, and are based on random number generating algorithms. A true random number is difficult to generate, but mathematically one can define tests of randomness. In a quantum computer, the measurement process is probabilistic, and the collapse of the wave function has origins in a non-deterministic process \cite{qrn,qrn1,test}. This has given rise to papers postulating random number generation using quantum computers \cite{test}. Independently, while trying to obtain quantum algorithms for non-linear differential equations, we observed that the measured output require a lot of post-processing to get accurate results \cite{lorenz,lorenz2}. Whereas the actual hardware noise can be improved with the building of fault-tolerant quantum computers, the inherent non-determinism of the measurement process is intrinsic to the quantum system. The probabilistic output, can be used to generate random numbers, as shown in \cite{qrn}. In this paper we show that {\it a non-linear ODE quantum simulation can be interpreted as a stochastic result}. Using qiskit and simulated noise-free circuits, we show how the stochastic Lorenz system, can be modeled using the measured quantum output. We compare the quantum generated trajectory for a deterministic Lorenz ODE with the example of the stochastic Lorenz model with classical randomness inbuilt into it. We conclude for this specific example that in a quantum computation code, the stochasticity emerges naturally in the measured outcomes of the quantum computer. The usefulness of quantum algorithms is in the promised efficiency and bigger processing power of scaled up quantum hardware. Thus quantum computers are viable alternate computational platforms for systems that do not require high precision, but bigger information storage and processing such as non-linear stochastic system. Whereas in this paper we show the equivalence of the quantum out put to the classical stochastic system for a specific system, the results can be tested for various other ODE's \cite{lorenz2}. In particular, the typical form of the finite difference iterative step matrix as modeled using a Unitary matrix yields $X_{i+1}= U X_{i}$, when measured from a quantum circuit, yields $\tilde{X}_{i+1}= X_{i+1} + \xi_i $, where $\xi_i$ is generated due to the finite shot measurements. These can be formulated as, and $ \xi_i \sim e\sqrt{h} N_i$ where $h$ is the time step and $N_i$ is a random number \cite{lorenz2}. A control parameter $e$ can be estimated as a function of the number of shots and the norm estimation \cite{lorenz2}.


The modeling of complex non-linear systems using deterministic chaos has been well studied \cite{lorenz}. These, when solved numerically, show regions of attractors and deterministic chaos. The stochasticity of these models have been studied using Euler-stochastic numerical equations \cite{l-84}. The behavior of these systems when driven by white noise facilitates synchronization \cite{synch}. In the following we discuss the time evolution of a stochastic Lorenz system.

{\bf The Lorenz System}\\
With the aim of observing stochastic behavior, we developed quantum algorithms for the Lorenz systems \cite{lorenz}. The Lorenz system, which we label as Lorenz-63, is a dynamical system in three variables $(x,y,z)$, representing thermodynamic quantities such as temperature and pressure gradients. There are three parameters, $(\sigma,\rho,\beta)$, which correspond to certain properties of the fluid/liquid.

The Euler-stochastic version \cite{synch} of the Lorenz system is
\begin{eqnarray}
    x_{i+1}&=& (1-h\sigma)\  x_i + \  h \sigma y_i \label{eqn:stoch1}, \\
    y_{i+1}&=& h \ \rho \  x_i - h \ x_i z_i +(1-h) y_i + e\sqrt{h}\ \xi_i \label{eqn:stoch2}, \\
    z_{i+1} &=& h\ x_i y_i + (1- h\beta) \ z_i .\label{eqn:stoch3}
\end{eqnarray}
Apart from the usual discretized variables ($x_i,y_i,z_i$, i=1..N) and `h' as the width of time step, $\xi_i$ is generated from a random number distribution. This represents the `random' degree of freedom in the Euler-Stochastic method \cite{stoch}, `$e$' is the strength of the stochastic term, following \cite{synch}. One can also add the stochastic terms to each of the three equations of the system. In each of the three cases the stochastic trajectories generated have very similar behavior \cite{lorenz2}. If we add different noise terms to the three equations simultaneously, the noise tends to overwhelm the system energetically.  We study the system in the region where the solutions are stable, so as to be able to compare the stochasticity generated by two different random number generators, one being classical, the other being the quantum measurement process.  This gives in general a bound on $e$, as in the above equation $e\leq O(1)$. Note when we obtain the output using a quantum circuit, the e can be fixed by comparison to the quantum circuit runs and has been estimated as $\sim (1/\sqrt{3 N})e_{i \ \rm norm}$, where $N$ is the number of shots and $e_i$ is the deviation of the norm from 1 in each time step. The details which make our classical stochasticity model give the quantum output are in \cite{lorenz2}. Our aim, however, is to eventually simulate new stochastic behaviors from quantum, not yet modeled using classical equations.

{\bf The Quantum Randomness}\\
In quantum mechanics, measurement is a probabilistic process. For a single qubit state $|q_0\rangle = \cos\theta |0\rangle + \sin\theta |1\rangle$, finite-shot sampling over $n$ shots yields frequencies that deviate from the exact probabilities $p_0=\cos^2\theta$ and $p_1=\sin^2\theta$. The number of successful outcomes follows a binomial distribution,
\begin{equation}
    P(r)= {}^nC_r p_0^r(1-p_0)^{n-r},
\end{equation}
with variance $\nu=np_0(1-p_0)$. The resulting statistical error in the estimation of $p_0$ is of the order of $\sqrt{p_0(1-p_0)/n}$, which means that the finite number of shots dictates the precision of the sampling. Generalizing to a multi-qubit circuit with a $d$-dimensional Hilbert space, Aaronson demonstrated that $O(d\log(d)/\epsilon^2)$ shots are required to achieve an $\epsilon$-precision characterization of the quantum state. Previous estimates had shown this as $O (d^2/\epsilon^2)$ using group theoretic analysis for a tensor product of $n$ copies of a $d$- dimensional density matrix \cite{donnelwright}. Thus, even without quantum noise, the precision with which the quantum state of a qubit is measured is not exact given the number of shots is finite. Whereas these deviations can be adjusted post-processing, we suggest an interesting use of the variance, which we label as {\it quantum finite shot randomness}. If we ask the question {\it how to generate} a random number using the quantum computer, the answer which the current understanding provides in the literature is the use of the wavefunction collapse in the measurement process \cite{test,test1}. These random number generation algorithms have survived the random number test \cite{test1}, and can be used genuinely for stochastic modeling. Our efforts to generate accurate modeling of the Lorenz system using quantum algorithms lead us to the same conclusion, that the output has a degree of randomness \cite{lorenz2}. Next, we show two different quantum algorithms for the Lorenz-63 system, both of which, in the measured output, show a degree of randomness which is stochastic. As predicted above, a simple output of a qubit measurement follows a binomial distribution with a variance. A two-qubit system with unentangled measurements represents a quadrinomial distribution, with variance in the measurement of all four quantum states. The use of binomial and multinomial distributions to generate random numbers is well known \cite{dist} and q-measurements can generate these. The quantum random number generator has use in the commercial world, and has been investigated for multiqubit systems in \cite{test1}, and here we suggest another use, that is, in modeling stochastic natural systems.

Next, we describe two quantum algorithms used to generate a Lorenz system, and in each, we report the quantum measured output as deviated from the numerically generated trajectory. Other methods to obtain solutions of non-linear ODE's, including discussions of the VQLS and post-processing methods, can be found in \cite{lorenz,lorenz2}. Note that the VQLS is a hybrid method, and the optimization procedure is post quantum measurement, and therefore has a post-processing method in place \cite{lorenz}. 

{\bf The S-FABLE Algorithm}\\
This method, which uses the existing repository of sparse-Fable (S-FABLE) coding of the numerical matrix equation, can be found in \cite{sfable}. The Lorenz numerical matrix equation comprises a variable vector $X= (x,y,z,xy,xz,1,1,1)^T$ which has non-linear terms as input and is motivated by a previous paper on solving the Lorenz system \cite{lorenz}. The matrix A, which represents the discrete equations (Eqs. \ref{eqn:stoch1}, \ref{eqn:stoch2} \& \ref{eqn:stoch3}) without the stochastic term, i.e. $e=0$ is:
\begin{equation}
    A= \left(\begin{array}{cccccccc}1-h\sigma&h\sigma&0&0&0&0&0&0\\h\rho&1-h&0&0&-h&0&0&0\\0&0&1-h\beta&h&0&0&0&0\end{array}\right)
\end{equation}

 To implement this operator, instead of directly encoding $A$, the matrix is first conjugated into the Walsh domain via $M = H^{\otimes n} A_{\text{scaled}} H^{\otimes n}$ where $H^{\otimes n}$ is the Walsh-Hadamard transform  and $A_{\text{scaled}}=S A S^{-1}$ is the version transformed by similarity of $A$ \cite{clader2013preconditioned, tong2021fast}, where $S$ is a diagonal rescaling matrix that divides each variable by its expected maximum range
$
    S = \mathrm{diag}\left(
\left[
\frac{1}{20},
\frac{1}{30},
\frac{1}{50},
\frac{1}{1000},
\frac{1}{600},
1,
1,
1
\right]
\right).$ This is done to prevent precision starvation under finite-shot sampling. The compilation represents the matrix elements as a sequence of single-qubit $R_y(\theta)$ rotations on the ancilla register, intercalated with CNOT gates ordered via a Gray code. The S-FABLE methodology then compresses this circuit by applying a simplification threshold $\epsilon_{\text{cutoff}} \approx 10^{-4}$ in the Walsh-Hadamard basis. Any rotation with an angle less than $\epsilon_{\text{cutoff}}$ is ruled out. The block-encoding of $A_{\text{scaled}}$ is then restored on the system register by applying Hadamards before and after the FABLE block:
\begin{equation}
U_{\text{S-FABLE}} = (I^{\otimes a} \otimes H^{\otimes n}) U_{\text{FABLE}} (I^{\otimes a} \otimes H^{\otimes n})
\end{equation}

yielding $\langle 0|^{\otimes a} U_{\text{S-FABLE}} |0\rangle^{\otimes a} = A_{\text{scaled}} / \alpha$, where $a = 4$ is the number of ancilla qubits, and $\alpha$ is the block-encoding scaling.

\begin{figure*}[]
    \centering
    \includegraphics[width=\linewidth]{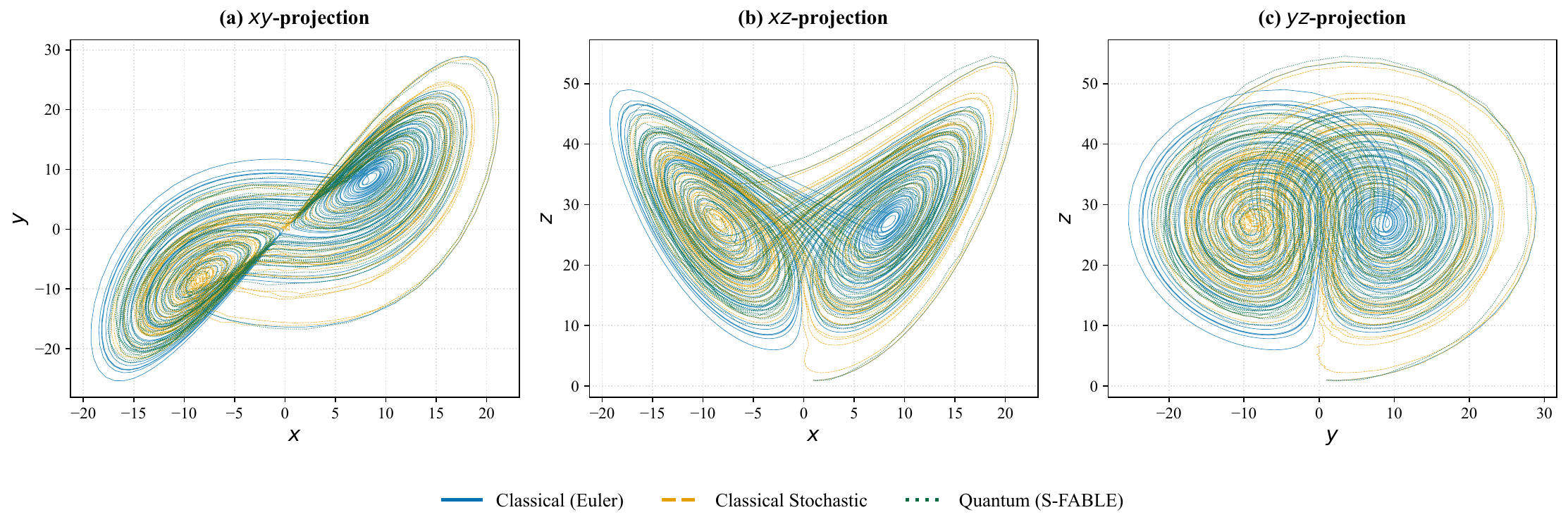}
    \caption{2D phase space projections of the Lorenz trajectories. Comparison between Classical Euler, Classical Stochastic and Quantum S-FABLE algorithm}
    \label{fig:lorenz2}
\end{figure*}

At each step, the block-encoding circuit $U_{\text{S-FABLE}}$ is executed on the initial state $\vert 0 \rangle^{\otimes a} \vert \psi_k \rangle$, where $\vert \psi_k \rangle$ is the normalized previous state. Taking the square root and substituting the normalization factors back, we can reconstruct the physical magnitudes $|X_j|$ at the new step and using the joint probability $p_{j,\text{joint}}$ of measuring the system in state $j$:$
|X_j| = \sqrt{p_{j, \text{joint}}} \cdot 2^n \cdot \|A_{\text{scaled}}\|_2 \cdot \|X_{\text{prev}}\|_2.$ To handle the loss of phase during measurement, signs are reconstructed using an Eulerian temporal continuity heuristic:
$\text{sign}(X_j) = \text{sign}(X_{j, \text{prev}} + \Delta t \cdot f_j(X_{\text{prev}})).$ The generated statevectors are measured using 500,000 shots, and the statevectors obtained using the estimates are plotted in the Figures (\ref{fig:lorenz2}). 

 Next we simulate stochastic behavior of the Lorenz system from a classical numerical simulation. Here, we show the evolution of the stochastic version of the Lorenz system using an Euler-Stochastic time evolution (Eqs. \ref{eqn:stoch1},\ref{eqn:stoch2} \& \ref{eqn:stoch3}). We take $\xi_i$ from a set of random numbers generated by Python's random number generator (which could be the binomial distribution). The result of the trajectories for using the same parameters as the S-FABLE coded trajectories are shown in (\ref{fig:lorenz2}).

To quantify the agreement between the quantum-measured trajectories and the classical stochastic Lorenz system, we test for the stochastic parameters of the two trajectories with 3 metrics: The Jensen-Shannon divergence between their power spectra is $D_{\mathrm{JS}}(Q,S)=[7.05,7.27,43.4]\times 10^{-3}$ along $x,y,z$, confirming that the quantum measurement noise reproduces the same spectral fingerprint as the classical stochastic system. The Pearson spectral correlation $R_{\mathrm{PSD}}(Q,S)$ exceeds $99.4\,\%$ in all components. The capacity and correlation fractal dimensions of the quantum trajectory, $D_0=0.844$ and $D_2=1.476$ (averaged over the three $2$D projections), are within $1\,\%$ of those of the stochastic counterpart. Detailed procedures and full results are reported in Appendix~\ref{app:2}.
\begin{figure}[htbp]
    \centering
      \includegraphics[width=0.8\linewidth]{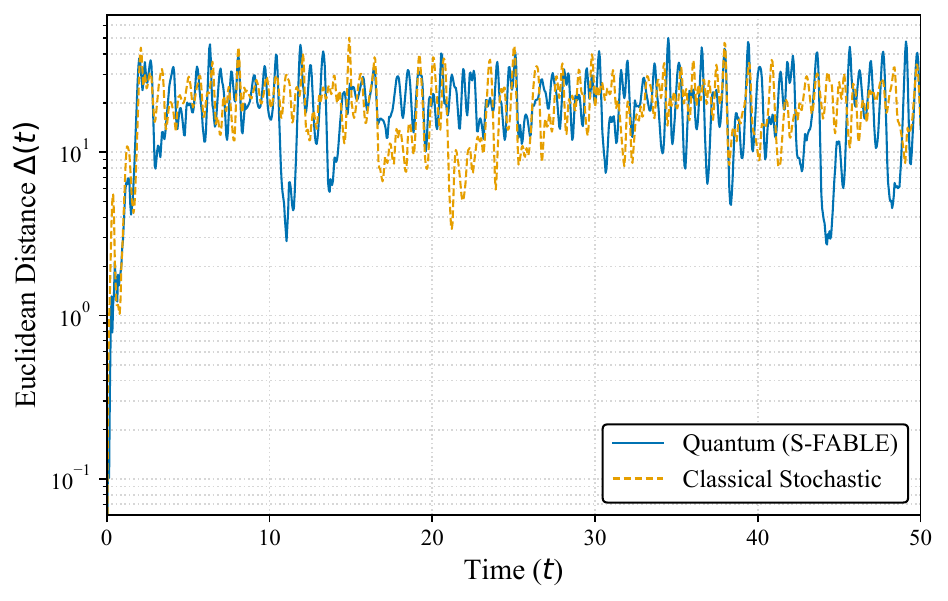}
        \caption{Euclidean distance error for Quantum S-FABLE algorithm and Classical Stochastic fluctuation relative to the classical Euler trajectory.}
        \label{fig:semi-log}  
\end{figure}
A plot of the classical stochastic deviations incurred, using the Euclidean distance between the stochastic and non-stochastic trajectories, is similar to the semi-log plot of that obtained by the quantum measurement of the S-FABLE circuit as in Figure (\ref{fig:semi-log}). In the classical system we have taken e=0.3 to reduce the fluctuations per time interval and the system is graphed. This shows the same pattern of the quantum fluctuations but using the semi-log scale.

\begin{figure}[htbp]
\centering
    \begin{subfigure}{0.4\textwidth}
    \centering
      \includegraphics[width=0.75\textwidth]{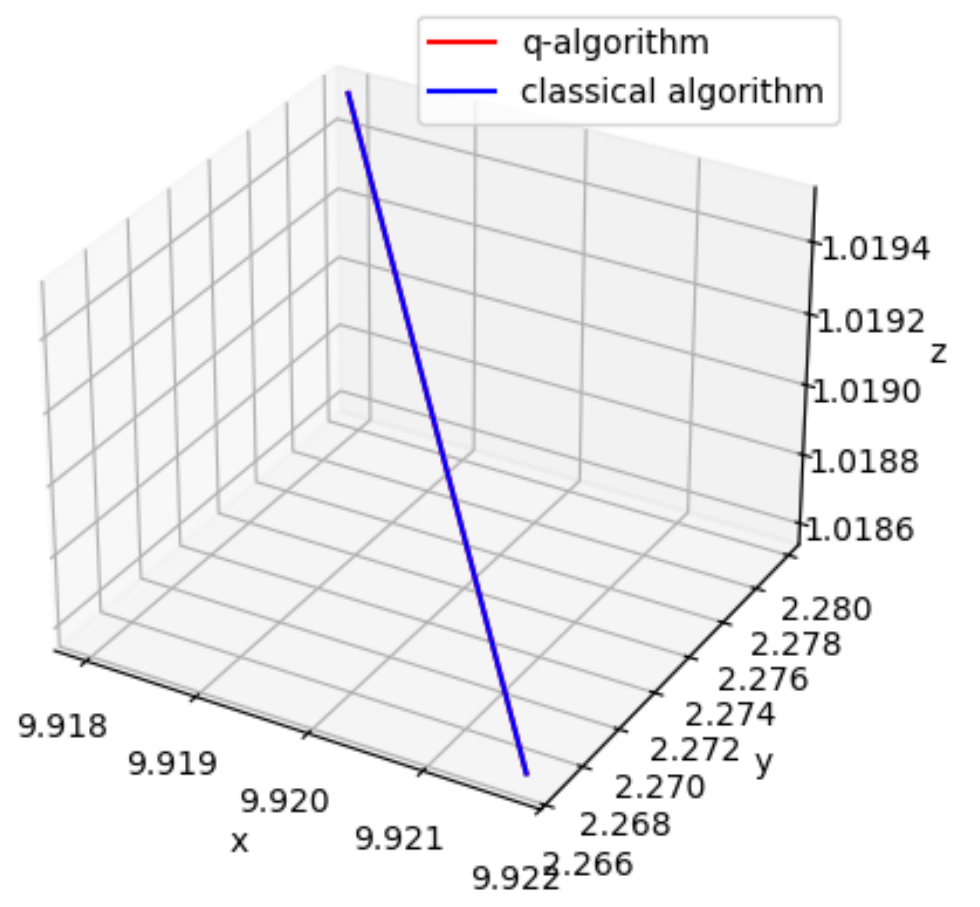}
      \caption{The sample of a trajectory starting at $x=10,y=2, z=1$ using state-vectors, for $\Delta t=0.00005$}
      \label{fig:lorun2}
    \end{subfigure}
    \hfill
    \begin{subfigure}{0.4\textwidth}
    \centering
      \includegraphics[width=0.75\textwidth]{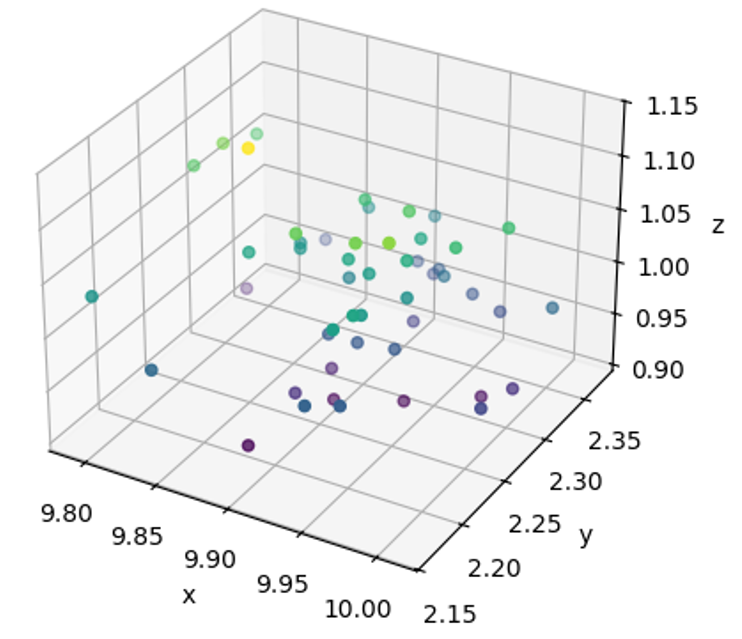}
      \caption{The sample of a trajectory starting at $x=10,y=2, z=1$ output obtained after measurement process $\Delta t=0.00005$}
      \label{fig:lorun3}
    \end{subfigure}
    \hfill
    \caption{The Quantum Algorithm generated trajectories}
    \end{figure}

\begin{figure}
 \centering
    \includegraphics[width=\linewidth]{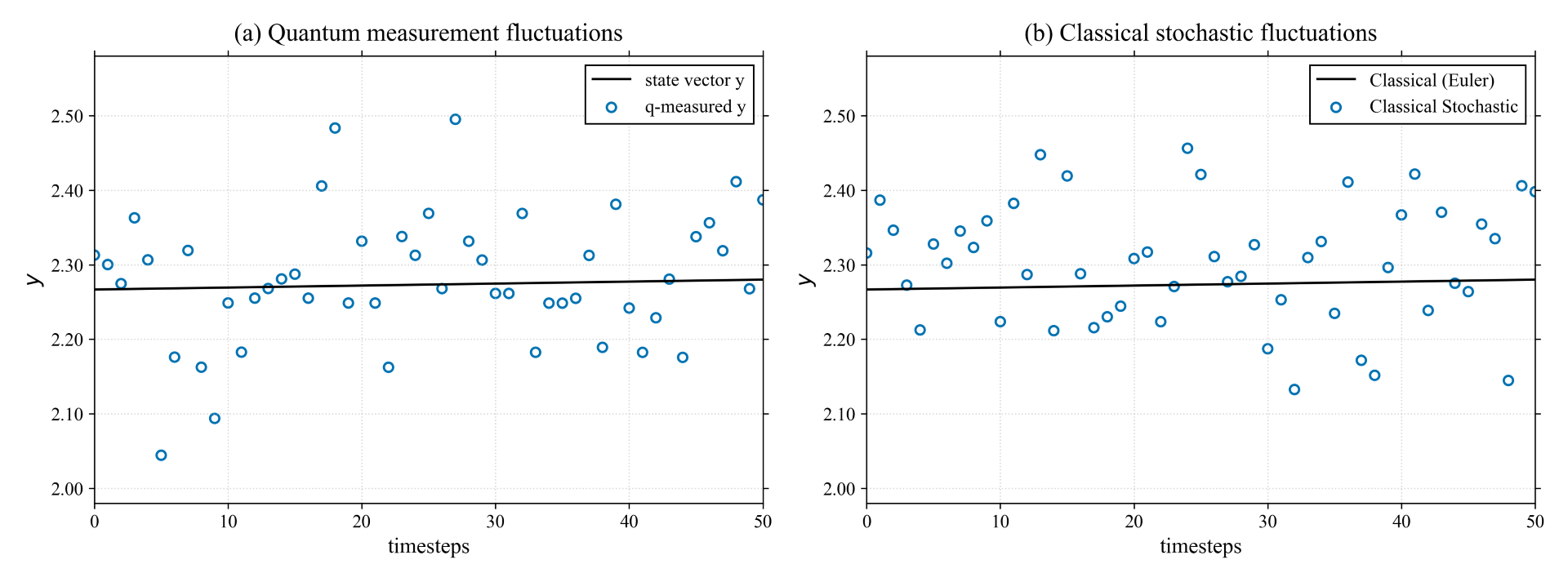}
    \caption{Comparison of stochastic fluctuations in the Lorenz system for y-component: \textbf{(a)} Quantum measurement distribution, \textbf{(b)} Numerically generated stochastic system.}
    \label{fig:yvalues}
\end{figure}

{\bf Unitary Evolution}\\
Instead of analyzing the randomness generated by the S-FABLE circuit, next, we describe a straightforward time evolution code, implemented using a Unitary matrix, which also has a block implementation. Further details of the algorithms and their use to solve ODE's can be found in \cite{lorenz2}. The Unitary circuit when measured can be modelled to generate a multinomial random distribution, as the entanglement of the qubits is minimal. {\it This `Unitary' matrix circuit for ODE is completely new, and easier to implement for the ODE systems.}
In the unitary algorithm, we time evolve the system using a Unitary matrix; however, it is different from the Schr\"odinger evolution. We write the discretized system using an antisymmetric (anti-Hermitian) matrix, and then exponentiate it to obtain a Unitary evolution. We use a  `transfer matrix' $A'$, such that $X_{i+1}=(I+h ~A') X_i$, and we build $Q$, 
$$Q= \left(\begin{array}{cc}0&A'\\-A'^{\dag}&0\end{array}\right)$$
such that $Q=-Q^{\dag}$. We then obtain a Unitary matrix $U=\exp(h Q)$. This method provides efficiency when one can implement the quantum algorithm using quantum parallelism \cite{lorenz2}. For this paper, we use this as an alternate quantum code to verify the stochastic nature of the Lorenz output. The matrix $A'$ for the Lorenz system is taken as 
\begin{equation}
 A'= \left(\begin{array}{cccccccc}-\sigma&\sigma&0&0&0&0&0&0\\\rho&-1&0&0&0&0&-1&0\\0&0&-\beta&0&0&1&0&0\end{array}\right)
 \label{eqn:matrix}
\end{equation}
The initial data is mapped onto the state vector $|X\rangle$ as:
\begin{equation}
    |X\rangle=\begin{pmatrix} 1 & x \end{pmatrix}^T \otimes \begin{pmatrix} x & y & z & 0 \end{pmatrix}^T = \begin{pmatrix}
    x & y & z & 0 & x^2 & xy & xz & 0\end{pmatrix}^T.
    \label{eqn:init}
\end{equation}
Though the minimal number of qubits to describe the Lorenz system is 3, due to the anti-Hermitian construction of $Q$, the time evolution is executed on a four-qubit circuit to accommodate the doubled Hilbert space dimensions, yielding the accurate trajectories shown in Figure (\ref{fig:lorun2}). Of the 16 state vector elements, only the first 3 are relevant and are retained in the code for the next iteration as in Equation(\ref{eqn:init}). We code with $h=10^{-6}$, and generate the solution for 50 consecutive time steps, iteratively, by repeated application of the Unitary matrix $X_{i+1}= U X_i$. After the quantum state is generated, a measurement process is made with the probabilities of $|000\rangle, |001\rangle, |010\rangle$, being recorded.
We focus on a small interval of width $\Delta t=[0.001-0.00105]$, with an initial condition of ($x=10,y=2,z=1$, at $t=0$) we generate the solution and evolve for $1000+i$ steps, where $i$ ranges from $0$ to $50$. The measurement on the quantum state is performed on the $1000+i$ th iteration. The error in the trajectory as read from the quantum algorithm's state vector is less than $10^{-6}$  as in Figure (\ref{fig:lorun2}). However, what we record is post measurement of the quantum state at the $1000+i$ th step, and use the square root of the probabilities to obtain the $x,y,z$ grid points. As the region of the trajectory is carefully chosen, all the gridpoints have positive values. We find that, the quantum measured output points are randomly distributed about the trajectory as in Figure (\ref{fig:lorun3}). The $y$ values extracted from the quantum state and from the measurement are shown in Figure(\ref{fig:yvalues} (a)), illustrating the stochastic nature of the quantum-measured results. To model our quantum output classically in this short interval, we have actually obtained a distributional `fit' for the quantum randomness of the circuit, and fed it into the classical equations. 

We generate a Python binomial random number of sample size $50$, with $10^6 $ trial shots and $p_0=0.068685$ and compare it with the measured $x$ values using the q-circuit. The details are given in the Appendix A. We find that both the samples of size 50 are same stochastically. We conclude that the quantum circuit we used to get the {\it Lorenz ODE trajectory can be used to generate multinomial distribution random numbers.}

It is evident that the quantum measured Lorenz output are stochastic as in Figure (\ref{fig:lorun3}) though the matrix we coded (\ref{eqn:matrix}) had $e=0$ in Equation (\ref{eqn:stoch1},\ref{eqn:stoch2},\ref{eqn:stoch3}).

A piece of the trajectory $t=[0.001-0.00105]$ using the same initial values (x=10,y=2,z=1) and the width of the time step $h=10^{-6}$ as in the quantum system , but using stochastic equations (\ref{eqn:stoch1},\ref{eqn:stoch2} \& \ref{eqn:stoch3}) with $e=1.62$, gives the same behaviour. To illustrate if one plots the y-values generated classically one gets exactly similar behavior in Figure (\ref{fig:yvalues} (b)) as its quantum measured output in (\ref{fig:yvalues} (a)) .

In this Unitary Matrix algorithm, we have directly taken the randomness of the quantum circuit as estimated by a binomial distribution, and simulated the classical stochastic system using the distribution. This is facilitated by the small number of time steps which allowed us to process the quantum output's nature of stochasticity. For bigger data, we will require more sophisticated methods of fitting the quantum output with random distributions.


We have shown that the quantum Lorenz trajectory has same behavior as that of a classical Lorenz trajectory with an additional stochastic term. Generalizing our result, we see that if we keep unprocessed output for a quantum ODE solver, the predicted trajectories will be stochastic. Although the practical utility of this approach depends on the system, it is promising for complex systems such as weather modeling, motivating our focus on the Lorenz system. 
Note that in this letter, we have studied the stochastic Lorenz system in a relatively stable region of the trajectories as in Figure (\ref{fig:lorenz2}); it will be interesting to observe the quantum output near bifurcation points. Regions of unstable behavior, particularly for complex systems far from equilibrium, would be interesting to simulate using quantum measurement and tomographic methods. In the quantum code labeled as Unitary evolution, we have also compared/modeled the measurement output using a multinomial distribution, which assumes that the qubits are unentangled. In conclusion, we observe that quantum circuits can be used to generate random numbers, and we can use that property to simulate classical stochastic systems. {\it All the codes used in this paper are available on request.}\\

{\bf Acknowledgment:} We would like to thank Robert Benkoczi, Daya Gaur, Shajad Hafshejani, Alfredo Iorio, Rossitsa Yalamova for useful discussions. Funding from ORIS, University of Lethbridge is acknowledged, and DC would like to thank MITACS. We are grateful to Urbasi Sinha for arxiv quant-ph endorsement. The authors have no conflict of interest to declare.

\appendix
\section{Random Number Sets}\label{app:1}
\begin{figure}
    \centering
      \includegraphics[height=1.5 in]{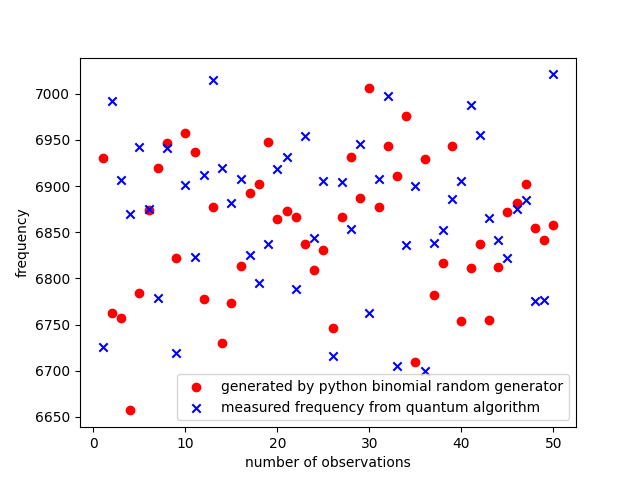}
      \caption{Random numbers generated by Python in red and the measured frequencies of $|000\rangle$ of the quantum state in blue.}
      \label{fig:random}
    \end{figure}

We took the sample for the $x$-data which was used to generate the $x$ values from the quantum measurement process for the Unitary matrix circuit. and calculated the mean and variance of the $50$ values generated using the measurement process. From the $50$ values of the frequency data for the $x$-values, the $mean/n$ is $0.068685$ and the $variance/n$ is $0.066397$. We observe that the sample size is $50$, and the above is drawn from a random set of numbers; we can fit a multinomial distribution. The error in variance is $3\%$, which can be attributed to any entanglement which we haven't accounted for, as the Unitary matrix requires when decomposed into two dimensional matrices. As illustrated in Figure (\ref{fig:random}), we generate a Python binomial random number of sample size $50$, with $10^6 $ trial shots and $p_0=0.068685$. This would generate the same set as for one variable x from a multi-nomial distribution, with the probability of its first variable being detected as $p_0$.
\hfill
    \hfill
\begin{multline*}
    {\rm BRND}=\\
    [6930, 6762, 6757, 6657, 6784, 6874, 6920, 6947, 6822, 6958,\\ 6937, 6778, 6877, 6730,6773,  6813, 6892, 6902, 6948, 6864,\\ 6873, 6867, 6837, 6809, 6831, 6746, 6867, 6932, 6887, 7006,\\ 6877, 6943,6911, 6976, 6709, 6929, 6782, 6817, 6943, 6754, \\ 6811, 6837,
 6755, 6812, 6872, 6882, 6902, 6855, 6842, 6858]
\end{multline*}

 \begin{multline*}
     {\rm QFR}=\\
     [6726, 6992, 6907, 6870, 6942, 6875, 6779,  6941, 6719,6901, \\6823, 6912, 7015, 6920, 6882, 6908, 6825, 6795, 6837, 6919 \\6932,  6788, 6954, 6844, 6905, 6716,  6904, 6853,  6946, 6762, \\ 6908, 6998, 6705,  6836,  6900, 6700, 6838, 6852,  6886, 6905,\\ 6988, 6955, 6865,6842, 6822,  6875,  6885, 6775, 6777,  7021]
 \end{multline*}
In the above, QFR is the measured frequency, and BRND is generated by the Python binomial random number generating procedure with $10^6$ trial shots and $p_0=0.068685$. The variance of this sample is about 3\% more than the QFR. Upto this difference, the quantum generated frequencies indeed are from a multinomial distribution, and therefore they can be used to generate multinomial random numbers. We performed a two sample T-statistics test and a KS test, and the P-value  (0.3070, 0.6435) were above the significance level, showing that the two samples represent the same distribution.  

\section{Details for stochastic metrics}\label{app:2}
\subsection{Jensen-Shannon divergence of the power spectrum}
The power spectral density (PSD) of each component ($x,y,z$) is computed via Welch's method with a Hamming window of $N_{\text{window}}=256$ points, $50\,\%$ overlap, and an FFT of $1024$ points. Each spectrum is smoothed with a Gaussian kernel ($\sigma=3$) and normalized to a probability distribution $p_i = P_i / \sum_j P_j$. The Jensen-Shannon divergence between two distributions $p$ and $q$ is $D_{\mathrm{JS}}(p,q)=\tfrac{1}{2}\bigl[S(p\|m)+S(q\|m)\bigr]$ where $m=\tfrac{p+q}{2}$ and $S(p\|m)=-\sum_i p_i\log(p_i/m_i)$ is the Kullback-Leibler divergence. $D_{\mathrm{JS}}=0$ indicates identical spectra, values approaching $1$ indicate maximal difference.

\subsection{Fractal dimensions}
The capacity dimension $D_0$ is computed via box-counting on the three two-dimensional projections ($x$-$y$, $x$-$z$, $y$-$z$) of the attractor. We overlay a grid of boxes of size $\epsilon$ spanning the phase-space extent and count the occupied boxes $N(\epsilon)$. The dimension is extracted from the linear relation $\log N(\epsilon) = -D_0 \log \epsilon$ over $\epsilon\in[0.05,3.0]$. The correlation dimension $D_2$ is obtained from the Grassberger-Procaccia algorithm~\cite{gp}, which estimates the scaling $C(r)\sim r^{D_2}$ of the correlation integral $C(r)=\frac{1}{N^2}\sum_{i\neq j} \Theta(r-\|x_i-x_j\|)$ over a range of radii $r$.

\subsection{Spectral correlation}
The Pearson correlation coefficient between the PSDs of the quantum and stochastic trajectories is computed over all frequency bins. For a pair of spectra $P_A(f)$ and $P_B(f)$,
\begin{equation}
    R_{\mathrm{PSD}}(A,B)=\frac{\sum_k (P_A(k)-\bar{P}_A)(P_B(k)-\bar{P}_B)}{\sqrt{\sum_k (P_A(k)-\bar{P}_A)^2 \sum_k (P_B(k)-\bar{P}_B)^2}}\,,
\end{equation}
where the sums run over the frequency bins of the Welch-periodogram. $R_{\mathrm{PSD}}=1$ corresponds to perfectly correlated spectra. Values exceeding $99.4\,\%$ indicate that the quantum measurement noise preserves the spectral structure of the classical stochastic driving. We find $R_{\mathrm{PSD}}(Q,S)=99.89\,\%$ ($x$), $99.91\,\%$ ($y$) and $99.41\,\%$ ($z$).
\begin{table}[htbp]
\centering
\caption{Fractal dimensions and relative error between quantum (Q) and stochastic (S) trajectories.}
\label{tab:fractal}
\begin{tabular}{lccc}
\hline
Projection & $D_0$ [Q] & $D_0$ [S] & Rel.\ error [\%] \\
\hline
$x$-$y$ & $0.904$ & $0.865$ & $4.55$ \\
$x$-$z$ & $0.875$ & $0.832$ & $5.15$ \\
$y$-$z$ & $0.752$ & $0.710$ & $5.89$ \\
\hline
Projection & $D_2$ [Q] & $D_2$ [S] & Rel.\ error [\%] \\
\hline
$x$-$y$ & $1.601$ & $1.603$ & $0.17$ \\
$x$-$z$ & $1.476$ & $1.481$ & $0.38$ \\
$y$-$z$ & $1.351$ & $1.364$ & $0.93$ \\
\hline
\end{tabular}
\end{table}

\begin{table}[htbp]
\centering
\caption{Jensen-Shannon divergence $D_{\mathrm{JS}}(Q,S)$ and spectral correlation $R_{\mathrm{PSD}}(Q,S)$ for each component.}
\label{tab:psd}
\begin{tabular}{lcc}
\hline
Component & $D_{\mathrm{JS}}(Q,S)$ & $R_{\mathrm{PSD}}(Q,S)$ \\
\hline
$x$ & $7.05\times 10^{-3}$ & $99.89\,\%$ \\
$y$ & $7.27\times 10^{-3}$ & $99.91\,\%$ \\
$z$ & $4.34\times 10^{-2}$ & $99.41\,\%$ \\
\hline
\end{tabular}
\end{table}

\bibliographystyle{unsrt}  
\bibliography{ref}

\end{document}